\documentclass[twocolumn]{aastex62}

\usepackage[caption=false]{subfig}
\usepackage{float}
\usepackage{booktabs}
\usepackage{multirow}
\graphicspath{{./}{figures/}}
\usepackage{xcolor}
\newcommand\hl[1]{%
  \bgroup
  \hskip0pt\color{black}
  #1%
  \egroup
}

\revised{\today}
\submitjournal{ApJL}
\shorttitle{Reflection and Emission of Habitable Exoplanets}
\shortauthors{Madden \& Kaltenegger}

\begin{document}

\title{High-resolution Spectra for a Wide Range of Habitable Zone Planets around Sun-like Stars}

\correspondingauthor{Jack Madden}
\email{jmadden@astro.cornell.edu}

\author[0000-0002-4701-7833]{Jack Madden}
\affil{Cornell University,
Astronomy and Space Sciences Building,
Ithaca, NY 14850, USA}
\affil{Carl Sagan Institute, Space Science Building 311,
Ithaca, NY 14850, USA}

\author{Lisa Kaltenegger}
\affil{Cornell University, 
Astronomy and Space Sciences Building,
Ithaca, NY 14850, USA}
\affil{Carl Sagan Institute, Space Science Building 311,
Ithaca, NY 14850, USA}

\begin{abstract} 
The search for life in the universe is currently focused on Earth-analog planets. However, we should be prepared to find a diversity of terrestrial exoplanets not only in terms of host star but also in terms of surface environment. \hl{Simulated high-resolution spectra of habitable planets covering a wide parameter space are essential in training retrieval tools, optimizing observing strategies, and interpreting upcoming observations. Ground-based extremely large telescopes like ELT, GMT, and TMT; and future space-based mission concepts like Origins, HabEx, and LUVOIR are designed to have the capability of characterizing a variety of potentially habitable worlds. Some of these telescopes will use high precision radial velocity techniques to obtain the required high-resolution spectra ($R\approx100,000$) needed to characterize potentially habitable exoplanets.}

Here we present a database of high-resolution \hl{(0.01 cm$^{-1}$)} reflection and emission spectra for simulated exoplanets with a wide range of surfaces, receiving similar irradiation as Earth around \hl{12 different host stars from F0 to K7.}

Depending on surface type and host star, we show differences in spectral feature strength as well as overall reflectance, emission\hl{, and star to planet contrast ratio of terrestrial planets in the Habitable zone of their host stars.} Accounting for the wavelength-dependent interaction of the stellar flux and the surface will help identify the best targets for upcoming spectral observations in the visible and infrared. 
 
All of our spectra \hl{and model profiles} are available online. 
\end{abstract}


\section{Introduction}
Currently, about 4000 extrasolar planets have been detected orbiting Main Sequence stars with dozens of terrestrial planets orbiting in their habitable zone (HZ) \citep{Kane2016, Johns2018, Berger2018}. The detected rocky exoplanets in the HZ show a wide variety of sizes and stellar hosts. For now, we are unable to characterize their atmospheres.

If our Solar System is any indication \cite[e.g.][]{Madden2018Abio,Krissansen2016,Cahoy2010,Lundock2009,Traub2003}, exoplanets should show a large diversity in composition and surface type. Therefore, it is important to model a wide range of surfaces and stellar hosts for rocky planets to expand the spectral database we will use to characterize planets and search for signs of life in their atmospheres.  

Direct observations that provide reflection and emission spectra of habitable zone exoplanets are critical to identifying signs of life on exoplanets \cite[e.g.][]{Kaltenegger2017, Scheieterman2018, Fujii2018} and should be within the capabilities of the next generation of ground-based telescopes like the extremely large telescopes (ELTs) \hl{and mission concepts such as Origins, HabEx, and LUVOIR} \cite[e.g.][]{Arney2018,Snellen2017}. \hl{Spectrographs on the Extremely Large Telescope (ELT) like HIRES (0.3-2.5$\mu m$) and METIS (3-19$\mu m$) are designed for a resolution of $R=100,000$ \citep{Ramsay2020}. Our database provides spectra modelled at 0.01$cm^{-1}$, which translates into a minimum resolution of $R=100,000$ from 0.4 to 10$\mu m$ and a minimum of $R=50,000$ from 10 to 20$\mu m$}

Here we present a high-resolution database of 360 reflection and emission spectra of Earth-like planets with diverse surfaces, which evolved in the HZ of a wide range of Sun-like host stars. Our spectra are based on the atmosphere models described in detail in \cite{Madden2020mnras}. 

This database enables us to explore which of these planets provide the strongest atmospheric features for overall characterization \hl{as well as signs of life. Biosignatures in this work represent disequilibrium atmospheric chemistry suggesting biotic sources, namely the biosignature pairs of O$_2$ and CH$_4$, and CH$_4$ and O$_3$ \citep{Lovelock1965,Lederberg1965}.}  

Our high-resolution spectra \hl{show the effects surfaces and host stars can have on the detectability of atmospheric features of habitable-zone planets and is a tool to prioritize promising targets in upcoming observations.}

Our spectra provide an important step in expanding the references used for optimizing upcoming observations, training retrieval algorithms as well as providing comparison model datasets to analyze future observations. \hl{In addition, studies show that high-resolution ($R\approx100,000$) exoplanet spectrum can be isolated from the combined star-planet spectrum, using the radial velocity difference between the two objects \citep{Snellen2015,Rodler2014,Brogi2014,Fischer2016,LopezMorales2019}. high-resolution spectra models of habitable atmospheres are important in refining this technique and may allow characterization of planets even if they can't be resolved.} 

\hl{Section 2 describes our models, section 3 presents our results, and section 4 discusses and summarizes our paper.}

Our high-resolution spectra are available online at \href{https://doi.org/10.5281/zenodo.3912065}{DOI: 10.5281/zenodo.3912065}.

\section{Methods}
\subsection{Planetary and Atmospheric Model}
The atmospheric composition of Earth-like planets depends on the outgassing rates, the irradiation from its host star, subsequent photochemistry, surface type, and cloud coverage. Here, we define `Earth-like' to refer to an Earth-radius and Earth-mass planet with similar outgassing rates to the modern Earth. Our spectra use planetary models generated using a coupled 1D climate and photochemistry model with wavelength-dependent albedo, described in detail in \cite{Madden2020mnras}.  

By incorporating wavelength-dependent reflection of surfaces and decoupling clouds from the surface reflection \cite{Madden2020mnras} explored the relationship between surface type and stellar type in the context of habitability. \cite{Madden2020mnras} found that surfaces with high variability across the visible and near-IR displayed a wide range of surface temperatures across star type. Surfaces like vegetation and sand showed the biggest change in surface temperature between cool K and hot F-stars while flatter overall albedo such as basalt, granite, coast, and seawater showed less change in surface temperature between star type. The surface temperature ranges for the different planet models are shown in Table \ref{tab:albedos}.

\begin{center}
\begin{table*}[t]
\begin{center}
\begin{tabular}{@{}cccc@{}}
\toprule
\multirow{3}{*}{\textbf{Surface}}  & \multirow{3}{*}{\textbf{Source}}   & \textbf{Temp. Range} & \textbf{$\Delta$Temp.}    \\
                          &                                             & \textbf{(K)}         & \textbf{(K)}               \\
                          &                                             & \textbf{(F0V-K7V)}   &              \\ \midrule
Basalt                    & ASTER Basalt: Solid: Basalt.H5      & 315.5-296.4                  & 19.1                  \\
Granite                   & ASTER Alkalic: Solid: Granite.H1    & 314.4-295.2                  & 19.2                  \\
Sand                      & ASTER Brown loamy fine: 87P3468     & 311.8-280.2                  & 31.6                  \\
Grass                     & ASTER Grass: Unknown             & 314.7-280.8                  & 33.9                  \\
Trees                     & ASTER Deciduous: Unknown              & 312.4-278.8                  & 33.6                  \\ \cmidrule(lr){2-2}
\multirow{2}{*}{Seawater} & USGS Open Ocean SW2 (0.2-2.4$\mu m$) & \multirow{2}{*}{326.4-304.7} & \multirow{2}{*}{21.7} \\
                          & ASTER Seawater: Liquid (2.4+$\mu m$) &                              &                       \\ \cmidrule(lr){2-2}
\multirow{2}{*}{Coast}    & USGS Coast SW1 (0.2-2.4$\mu m$)      & \multirow{2}{*}{326.6-303.9} & \multirow{2}{*}{22.7} \\
                          & ASTER Seawater: Liquid (2.4+$\mu m$) &                              &                       \\ \cmidrule(lr){2-2}
Cloud                     & Modis $20\mu m$ Cloud Model          & 249.9-260.0                  & -10.1                 \\ \midrule
Basalt+Cloud              & 56.3\% Basalt, 43.7\% Cloud                 & 286.7-281.9                  & 4.8                   \\
Granite+Cloud             & 56.3\% Granite, 43.7\% Cloud                & 286.1-280.8                  & 5.3                   \\
Sand+Cloud                & 56.3\% Sand, 43.7\% Cloud                   & 284.0-271.9                  & 12.1                  \\
Grass+Cloud               & 56.3\% Grass, 43.7\% Cloud                  & 285.0-272.7                  & 13.8                  \\
Trees+Cloud               & 56.3\% Trees, 43.7\% Cloud                  & 283.9-270.1                  & 12.3                  \\
Seawater+Cloud            & 56.3\% Seawater, 43.7\% Cloud               & 297.0-287.8                  & 9.2                   \\
Coast+Cloud               & 56.3\% Coast, 43.7\% Cloud                  & 297.1-287.9                  & 9.2                   \\ \midrule
Basalt+Seawater           & 30\% Basalt, 70\% Seawater                  & 323.1-302.7                  & 20.4                  \\
Granite+Seawater          & 30\% Granite, 70\% Seawater                 & 322.9-302.1                  & 20.8                  \\
Sand+Seawater             & 30\% Sand, 70\% Seawater                    & 322.5-299.2                  & 23.3                  \\
Grass+Seawater            & 30\% Grass, 70\% Seawater                   & 323.2-299.0                  & 24.1                  \\
Trees+Seawater            & 30\% Trees, 70\% Seawater                   & 322.7-298.6                  & 24.2                  \\
Snow+Seawater             & 30\% Snow, 70\% Seawater                    & 290.6-288.5                  & 2.1                   \\ \midrule
Basalt+Seawater+Cloud     & 56.3\% (Basalt+Seawater), 43.7\% Cloud      & 293.7-286.1                  & 7.6                   \\
Granite+Seawater+Cloud    & 56.3\% (Granite+Seawater), 43.7\% Cloud     & 293.5-285.9                  & 7.6                   \\
Sand+Seawater+Cloud       & 56.3\% (Sand+Seawater), 43.7\% Cloud        & 292.6-283.1                  & 9.5                   \\
Grass+Seawater+Cloud      & 56.3\% (Grass+Seawater), 43.7\% Cloud       & 293.1-283.4                  & 9.5                   \\
Trees+Seawater+Cloud      & 56.3\% (Trees+Seawater), 43.7\% Cloud       & 292.6-283.1                  & 9.7                   \\
Snow+Seawater+Cloud       & 56.3\% (Snow+Seawater), 43.7\% Cloud        & 277.9-277.2                  & 0.7                   \\ \midrule
\multirow{3}{*}{Earth}    & 70\% Seawater, 2\% Coast, 2.52\% Basalt,    & \multirow{3}{*}{319.6-298.6} & \multirow{3}{*}{21.0} \\
                          & 2.52\% Granite, 1.96\% Sand, 8.4\% Grass,   &                              &                       \\
                          & 8.4\% Trees, 4.2\% Snow                     &                              &                       \\ \cmidrule(lr){2-2}
Earth+Cloud               & 56.3\% Earth, 43.7\% Cloud                  & 290.4-283.1                  & 7.3                   \\ \midrule
Flat                      & Flat reflectence of 0.31                    & 291.3-280.8                  & 10.5                  \\ \bottomrule
\end{tabular}
\end{center}
\caption{The 30 simulated surface types with source and surface temperature range across star types. USGS: \cite{Kokaly2017usgs} (https://crustal.usgs.gov/speclab/), ASTER: \cite{Baldridge2009aster} (https://speclib.jpl.nasa.gov/library), Modis: \cite{King1997}}
\label{tab:albedos}
\end{table*}
\end{center}

\subsection{Generating reflection and emission spectra}
We use EXO-Prime2 to generate the high-resolution reflection and emission spectra for each simulated exoplanet from 0.4 to 20$\mu m$ at a resolution of $0.01 cm^{-1}$. The radiative transfer model used was originally developed for stratospheric measurements in Earth's atmosphere \citep{TraubStier1976, TraubJucks2002} and has been updated for use with exoplanets \citep[e.g.][]{DesMarais2002,TraubJucks2002,Kaltenegger2007,KalteneggerTraub2009, OMalleyJames2019}. 
For our calculations, we used 38 plane-parallel layers for an 80km atmosphere with an observation zenith angle of 60 degrees giving an approximation of quadrature viewing. 

We include the molecular species with prominent absorption features expected in the atmospheres of Earth-like planets orbiting F to K stars as modeled in \cite{Madden2020mnras}. We use the 2016 HITRAN database for our opacities for H$_2$O, CO$_2$, CH$_4$, N$_2$O, O$_3$, O$_2$, H$_2$CO, OH, C$_2$H$_6$, HO$_2$, CO, NO, NO$_2$, H$_2$O$_2$, H$_2$S, and SO$_2$ \citep{Gordon2017hitran}. We include CO$_2$ line mixing \citep{Niro2005}. For CO$_2$, H$_2$O, and N$_2$, we use measured continua data instead of line-by-line calculations in the far wings \citep{TraubJucks2002}.

\hl{With no clear answer on how cloud-feedback should affect clouds on exoplanets orbiting different host stars, we use Earth's clouds as a first approximation for all our models. We include 3 cloud layers in our models (following \cite{Kaltenegger2007}) at 1km (40\%), 6km (40\%), and 12km (20\%) and an overall cloud coverage of 44\% \citep{Madden2020mnras}. This simulates an observation of a cloudy exoplanet by having the spectrum represent the sum of different layers in the atmosphere.}
\pagebreak
\subsection{Stellar Spectra \& Surface Albedos}
The effects of wavelength-dependent surface and cloud albedo on habitability are most apparent when comparing the planetary models across star type. We used the same ATLAS model \citep{CastelliKurucz2004} F, G, and K star as in \cite{Madden2020mnras} for our calculations of planetary spectra and star to planet contrast. In total, we simulated planets around 12 star types spaced roughly 250K in temperature between an F0V (7,400K) and a K7V (3,900K). \hl{Note that the simulations in \cite{Madden2020mnras} used lower total stellar incident flux on the planet for cooler host stars to achieve temperatures similar to modern Earth across star type ($288 K \pm 2$ percent).} 

The albedos used here and in \cite{Madden2020mnras} focus on the dominant surfaces on Earth: seawater, coastal water, basalt, granite, sand, trees, grass, snow, and clouds. A modern Earth albedo can be made by combining these surfaces with weights based on their modern Earth surface coverage \citep{Kaltenegger2007}. Surface albedos were taken from the USGS and ASTER spectral libraries ~\citep{Baldridge2009aster,Kokaly2017usgs,Clark2007usgs}. For \hl{all three} cloud layers, we use the $20\mu m$ Modis cloud albedo ~\citep{King1997,RossowSchiffer1999} (Table \ref{tab:albedos}). 

In this paper we show four planetary scenarios for each surface: i) a single planetary surface to show the maximum effect of a specific surface on the spectra, ii) a 30\% single surface and 70\% seawater combination, and iii) and iv) two more scenarios where these cases have the 44\% cloud coverage, derived to simulate the modern Earth model in \cite{Madden2020mnras}. 

For 30 different surfaces around 12 host star types, we simulated in total, 360 terrestrial planetary spectra from 0.4 to 20$\mu m$.

\section{Results}
Our high-resolution spectra database \hl{contains the combined reflection and emission spectra for 360 Earth-like planets with 30 different surfaces orbiting 12 different Sun-like host stars. All spectra shown are a combination of planet reflection and emission. Emission begins to dominate the flux between 3 and 4$\mu m$ depending on the star and surface temperature.} 

We highlight a subset of these spectra in our figures to show a balance of variety and specific effects while keeping figures uncrowded. We do not show the simulated spectra for granite, grass, coast, or cloud surface only. The surface reflectivity of granite is similar to first-order with basalt, grass with trees, and coast with seawater. However, these spectra can all be downloaded from our database. 

\subsection{Reflection Spectra}
The star-surface interaction leads to drastic differences in a planet's appearance, which are most apparent in the exoplanets' reflectance spectra in the visible. 

\begin{figure*}[ht!]
\centering
\includegraphics[width=\textwidth]{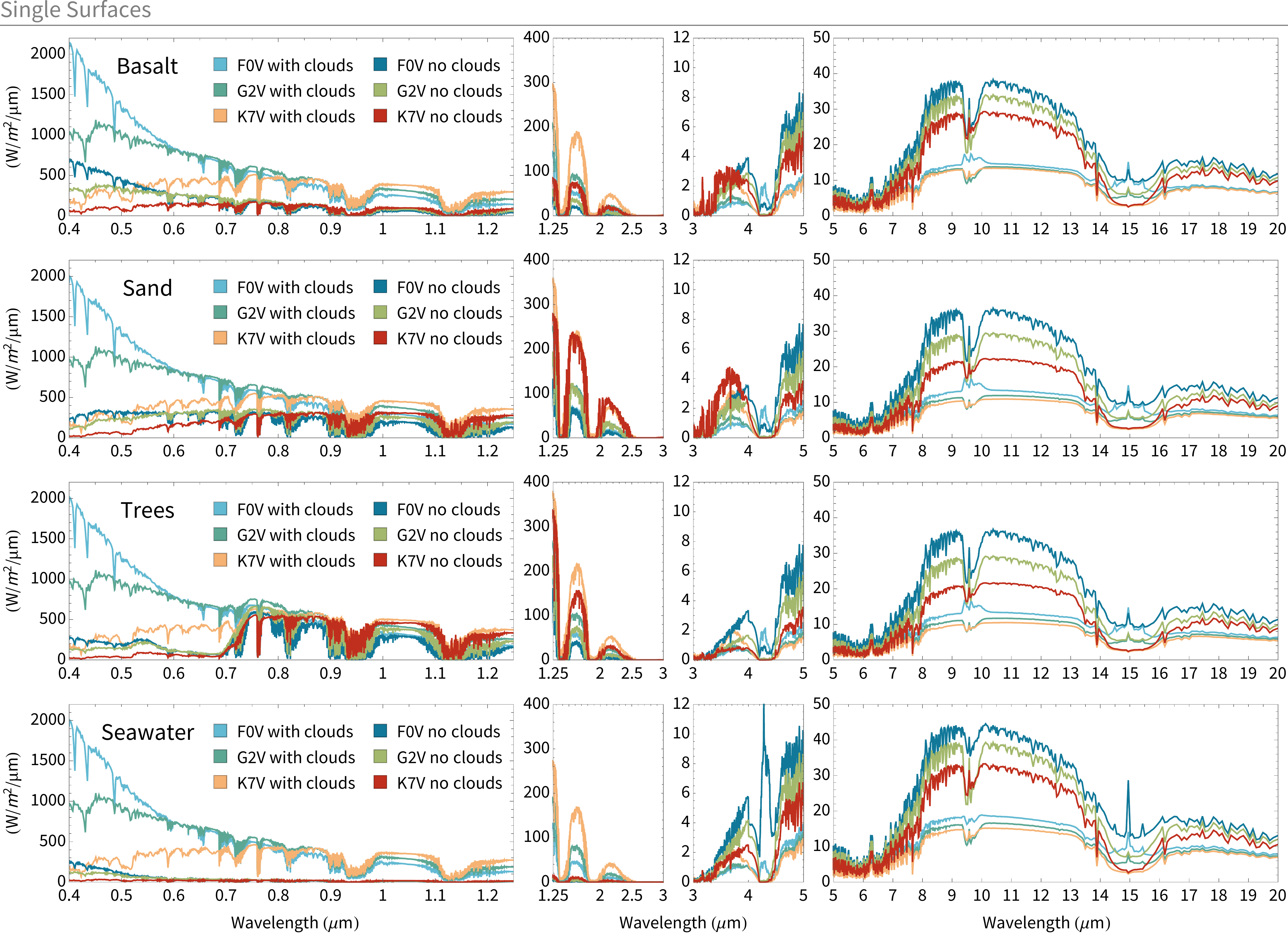}
\caption{A sample of the \hl{combined} reflection and emission spectra from the simulated exoplanets with 100\% of a single surface type both with and without clouds added. \label{fig:single}}
\end{figure*}

\begin{figure*}[ht!]
\centering
\includegraphics[width=\textwidth]{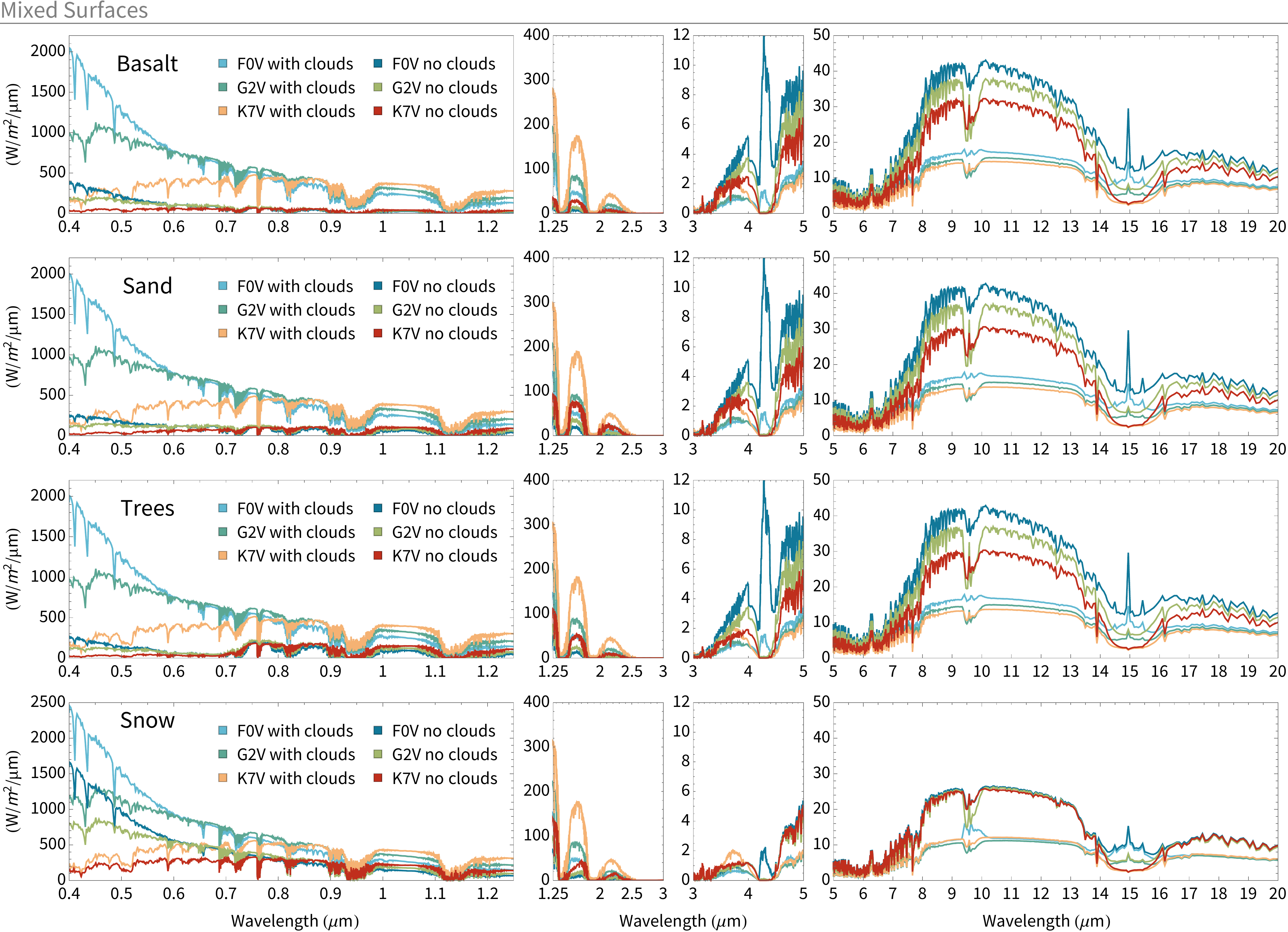} 
\caption{A sample of the \hl{combined} reflection and emission spectra from the simulated exoplanets with mixed surfaces of 30\% of one type and 70\% of seawater both with and without clouds added. \label{fig:mixed}}
\end{figure*}

\hl{Even though the incident stellar flux decreases for cooler star types to provide similar surface temperatures in our models, the reflected flux of a planet can vary by more than an order of magnitude at specific wavelengths depending on a planet's surface reflectivity, as shown in Fig. \ref{fig:single}, and Fig. \ref{fig:mixed}. It can result in planets with highly reflective surfaces orbiting cooler stars reflecting more starlight than planets with low surface reflectivity orbiting hotter stars: For example, at $0.5 \mu m$, a desert planet (sand surface) orbiting a K7V star is twice as bright as an ocean planet (seawater surface) orbiting an F0V host star, despite the higher incident flux of an F0V star at that wavelength.}

\hl{Surfaces with high reflectivity generally lead to more prominent spectral features at visible wavelengths. The deepest absorption features can be seen for planetary models with high reflective surfaces like vegetation, sand, and snow orbiting the hottest grid host stars, which provide the highest incident flux (Fig. \ref{fig:single} and \ref{fig:mixed}).} 

The shape of the surface albedo also modulates the flux of the visible exoplanet spectra models. For example, the vegetation 'red-edge' near $0.7 \mu m$ shows as a strong increase in reflectivity in the spectra of tree-covered planets (Fig. \ref{fig:single} and Fig. \ref{fig:mixed}).

\subsection{Emission}
Planetary surface albedos can have a large effect on the planetary surface temperature as well as atmospheric temperature structure \citep{Madden2020mnras}. \hl{Planetary models with highly reflective surfaces generally lead to lower surface temperatures and therefore lower infrared emission, while models with less reflective surfaces lead to higher surface temperatures and thus higher infrared emission of the planet for a specific host star} (Table \ref{tab:albedos}, Fig. \ref{fig:single} and \ref{fig:mixed}).

\hl{The models in \cite{Madden2020mnras} used a reduced incident flux for cooler stars to achieve similar modern Earth temperatures for a constant surface albedo of 0.31. Therefore, while the surface reflectivity of a planet changes the surface temperature for similar incident flux, that difference has been compensated for in the modeling for a specific wavelength-independent surface albedo case, resulting in a slight increase in surface temperature and overall emission for planets orbiting hotter host stars (see Fig. \ref{fig:single} and \ref{fig:mixed}).}

\hl{Infrared spectral feature depth depends on both the abundance as well as the difference in temperature of the overall emitting and absorption layer. Thus the deepest absorption features are not seen for the hottest planetary models with low reflective surfaces like oceans (Fig. \ref{fig:single} and \ref{fig:mixed}), because of the similarity in temperature of the two layers compared to planets with different surfaces.}

\subsection{Planet-to-star contrast}
\begin{figure*}[ht!]
\centering
\includegraphics[width=\textwidth]{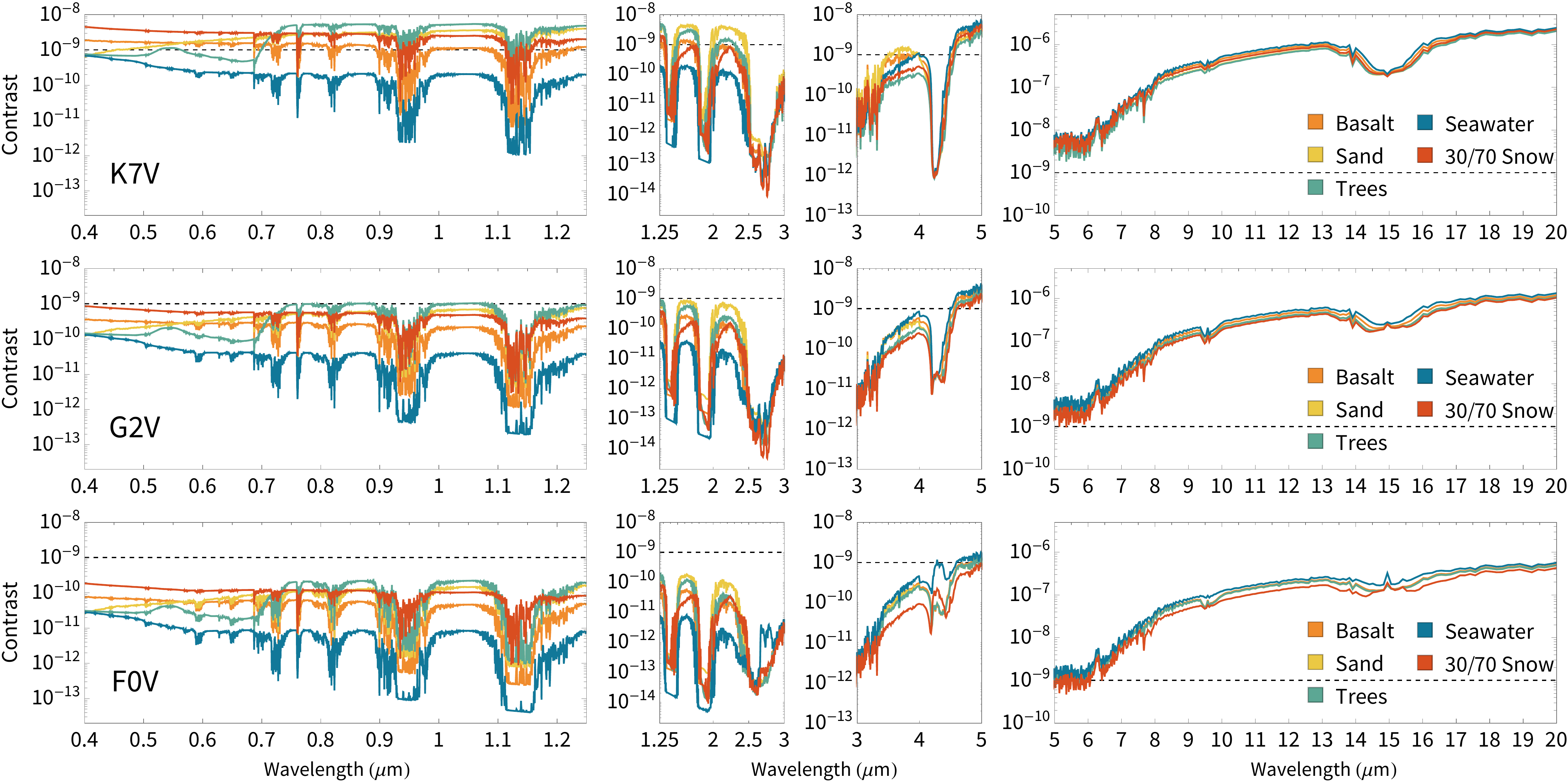}
\caption{Contrast for planets modeled with basalt, sand, tree, seawater, and snow surfaces around K7V (top), G2V (middle), and F0V stars (bottom). \hl{A line at $10^{-9}$ is shown for reference between panels.}  \label{fig:contrast}}
\end{figure*}

Planet-star contrast is generally higher for similar planets around cooler stars versus hotter stars. Fig. \ref{fig:Earth} shows that Earth-like planets orbiting our coolest grid stars have the highest contrast across the spectrum compared to hotter host stars. Planets with the same surface also show this based on our spectra (Fig. \ref{fig:contrast}). 

However, an ocean planet covered with dark seawater orbiting a K7V-star shows a similar contrast ratio in the visible and near-infrared ($0.7-4\mu m$) as a planet covered in vegetation around an F0V host star (Fig. \ref{fig:contrast}). Planets covered by highly reflective surfaces such as grass, trees, snow, and sand around G-stars will be consistently as high or higher in contrast at visible wavelengths than planets covered by darker surfaces such as coast, seawater, basalt, or granite around K-stars. Therefore, a planet’s surface can influence the contrast ratio significantly in the visible and near-IR. When comparing surfaces with extreme differences in reflectivity, cooler stars may not always provide the highest contrast habitable zone targets in the visible and near-IR, depending on their surface composition.

\subsection{Atmospheric composition change with host star}
\begin{figure*}[ht!]
\centering
\includegraphics[width=\textwidth]{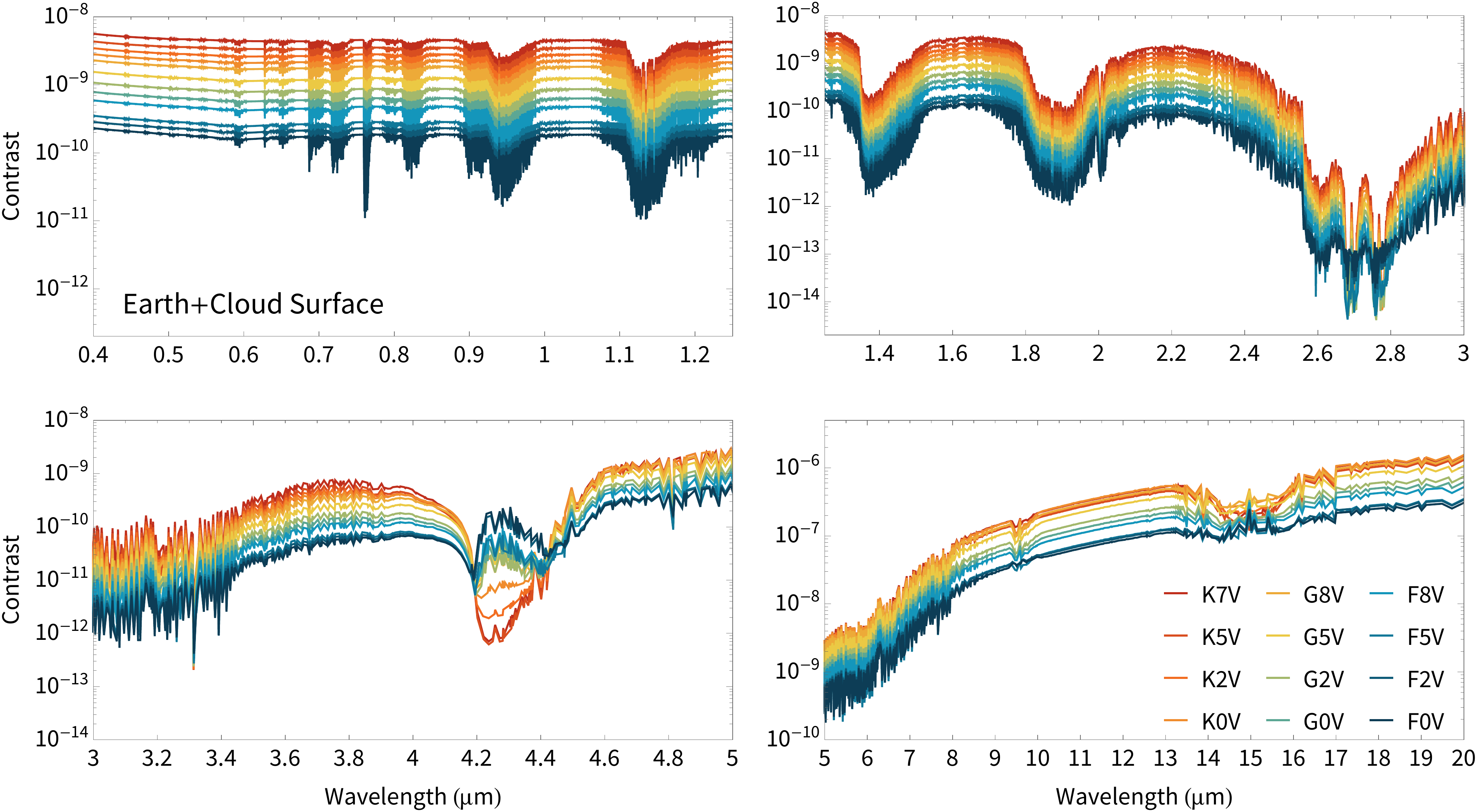}
\caption{The model contrast spectra for a modern Earth surface including clouds across all star types. \label{fig:Earth}}
\end{figure*}

The stellar energy distribution (SED) of a star influences the atmospheric composition of a planet \citep{Kasting1993,Rugheimer2013,Rugheimer2015Spectra,Segura2003,Segura2005,Madden2020mnras,Rauer2011}. \hl{Fig. \ref{fig:Earth}, Fig. \ref{fig:single}, and Fig. \ref{fig:mixed} show the varying depth of the atmospheric spectral features such as O$_3$, CO$_2$, and CH$_4$ for different stellar hosts.} 

In the visible, the depth of an absorption feature is proportional to the abundance of a molecule, the amount of incident stellar radiation, and the reflectivity of the planet. For similar reflectivity, the change in the absorption features depth reflects the change in abundance of the chemicals due to the stellar SED and subsequent photochemistry in the planet's atmosphere as well as incident irradiation (Fig. \ref{fig:Earth}). Hotter stars in our grid emit higher UV flux, thus altering the profiles of these molecules and subsequent reactions in a planet's atmosphere

In the infrared, the depth of the absorption features depends on the abundance of a chemical as well as the temperature difference between the emitting/absorbing layer and the continuum.

Surfaces can modify the atmosphere composition based on how the surface albedo alters the surface temperature of the planet as well as the temperature profiles of the atmosphere, for example how much water is evaporated. 

Fig. \ref{fig:Earth} shows the change in planet-to-star contrast ratio for a planet model with a modern Earth-analog surface for our grid stars to isolate the effect of the host star on the planet's spectra.

The most notable spectral features between 0.4 and 20$\mu m$ relevant to biosignature detection include oxygen and ozone at 0.69, 0.76, and 9.6$\mu m$; in combination with methane at 0.88, 1.04, 2.3, 3.3, and 7.66$\mu m$; ; N$_2$O shows features at 7.75, 8.52, 10.65, and 16.89$\mu m$. H$_2$O has features at 0.6, 0.65. 0.73, 0.82, 0.95, 1.14, 1.4, 1.85, 2.5-3.5, 3.7, and 5-8$\mu m$. Another greenhouse gas seen in the spectra is CO$_2$ at 2.7, 4.25, and 15$\mu m$.

\begin{figure*}[ht!]
\centering
\includegraphics[width=0.9\textwidth]{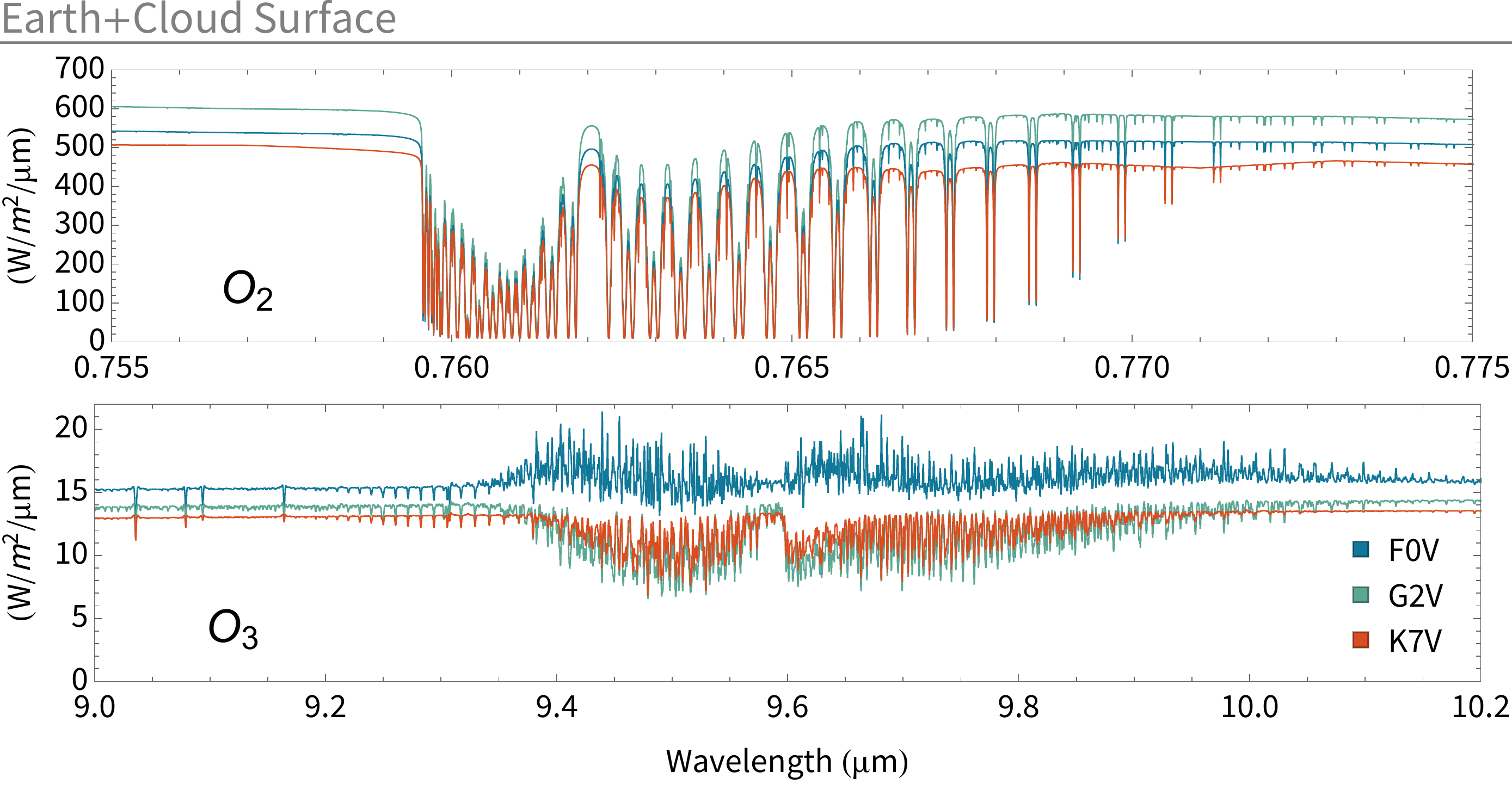} 
\caption{\hl{Two oxygen features shown at high-resolution ($0.01cm^{-1}$, $R>100,000$) for an exoplanet with an Earth-like albedo with clouds around an F0V, G2V, and K7V host star.}  \label{fig:highres}}
\end{figure*}

\hl{As an example, for the 0.76$\mu m$ O$_2$ and the 9.6$\mu m$ O$_3$ feature, Fig. \ref{fig:highres} shows that at a high-resolution of $0.01cm^{-1}$, all spectral features have a distinct series of lines to identify such Doppler-shifted lines uniquely on exoplanets which move predictably around their host star \citep{Rodler2014,Snellen2015,Brogi2014}. For specific absorption features of interest, our high-resolution spectra database can be used to optimize observation strategies for specific features and specific wavelength regions.}

\newpage

\section{Discussion \& Conclusion}
We present 360 emission and reflection spectra at a high-resolution of $0.01cm^{-1}$ for habitable zone planets orbiting 12 F, G, and K star types. These model spectra show the interaction of the host star's stellar energy distribution and a planet's wavelength-dependent albedo. \hl{We used the 9 dominant surfaces on modern Earth and isolated the effects of rock (basalt and granite), vegetation, sand, snow, clouds, water (ocean and coast) as well as clouds on a planet's reflection and emission spectra from the visible to the infrared (0.4-20$\mu m$).} 

\hl{To show the variety in this database, we include models with single surface planets to show the maximum effect of each surface on the spectra as well as the planet star to planet contrast ratio (Fig. \ref{fig:contrast} and \ref{fig:single}), mixtures of seawater and cloud coverage (Fig. \ref{fig:single} and \ref{fig:mixed}) as well as modern Earth surface coverage models (Fig. \ref{fig:Earth}) for the 12 host stars from F0V to K7V.} 

Other surface albedos and fractional combinations of surfaces are of course possible for exoplanets. As a first-order estimate our single surface spectra can be combined to create spectra for new surface mixtures. In this way, our spectral database provides a toolkit to generate estimated spectra of Earth-like planets with different surface combinations with and without clouds. One aspect that won't be captured by such a combination is the potential difference in surface temperature introduced by the different surface reflectivity \cite[see][]{Madden2020mnras}. Exoplanets may also have many different surface types not addressed in this study including mineral surfaces\citep{Shields2018}, a wide range of different biota \citep{Hegde2013} or biofluorescent organisms\citep{OMalleyJames2018fluor}.

We've shown that surface albedos can deepen spectral features. In the visible, where the spectra show reflected starlight, highly reflective surfaces generally increase the depth of absorption features. In the infrared, on the other hand, low reflective surfaces increase the surface temperature and thus a planet's overall emission. 

\hl{Our high-resolution spectra database provides a critical tool in the planning and analysis of observations with upcoming ground-based telescopes like ELT, GMT, and TMT and future space mission concepts Origins, HabEx, and LUVOIR. Ground-based telescopes plan to employ high precision radial velocity techniques that require high-resolution ($R>100,000$) in order to characterize potentially habitable exoplanets.}

Studying the wide range of changes caused by different surfaces and host stars improves our understanding of biosignatures and their remote observability. \hl{Obtaining high-resolution spectra of terrestrial planets in the habitable zone is an essential milestone in discovering life beyond our Solar System. Our database is intended to support this objective by presenting a wide range of cases for further study, planning, training, and comparison.}

\acknowledgments
This work was supported by the Carl Sagan Institute and the Brinson Foundation.
\pagebreak

\begin{thebibliography}{}
\expandafter\ifx\csname natexlab\endcsname\relax\def\natexlab#1{#1}\fi
\providecommand{\url}[1]{\href{#1}{#1}}

\bibitem[{{Arney} {et~al.}(2018){Arney}, {Batalha}, {Cowan}, {Domagal-Goldman},
  {Dressing}, {Fujii}, {Kopparapu}, {Lincowski}, {Lopez}, {Lustig-Yaeger}, \&
  {Youngblood}}]{Arney2018}
{Arney}, G., {Batalha}, N., {Cowan}, N., {et~al.} 2018, arXiv e-prints,
  arXiv:1803.02926

\bibitem[{Baldridge {et~al.}(2009)Baldridge, Hook, Grove, \&
  Rivera}]{Baldridge2009aster}
Baldridge, A.~M., Hook, S., Grove, C., \& Rivera, G. 2009, Remote Sensing of
  Environment, 113, 711

\bibitem[{{Berger} {et~al.}(2018){Berger}, {Huber}, {Gaidos}, \& {van
  Saders}}]{Berger2018}
{Berger}, T.~A., {Huber}, D., {Gaidos}, E., \& {van Saders}, J.~L. 2018, \apj,
  866, 99

\bibitem[{{Brogi} {et~al.}(2014){Brogi}, {de Kok}, {Birkby}, {Schwarz}, \&
  {Snellen}}]{Brogi2014}
{Brogi}, M., {de Kok}, R.~J., {Birkby}, J.~L., {Schwarz}, H., \& {Snellen},
  I.~A.~G. 2014, \aap, 565, A124

\bibitem[{Cahoy {et~al.}(2010)Cahoy, Marley, \& Fortney}]{Cahoy2010}
Cahoy, K.~L., Marley, M.~S., \& Fortney, J.~J. 2010, ApJ, 724, 189

\bibitem[{{Castelli} \& {Kurucz}(2004)}]{CastelliKurucz2004}
{Castelli}, F., \& {Kurucz}, R.~L. 2004, ArXiv Astrophysics e-prints,
  astro-ph/0405087

\bibitem[{Clark {et~al.}(2007)Clark, Swayze, Wise, Livo, Hoefen, Kokaly, \&
  Sutley}]{Clark2007usgs}
Clark, R.~N., Swayze, G.~A., Wise, R.~A., {et~al.} 2007, USGS digital spectral
  library splib06a, Tech. rep., US Geological Survey

\bibitem[{{Des Marais} {et~al.}(2002){Des Marais}, {Harwit}, {Jucks},
  {Kasting}, {Lin}, {Lunine}, {Schneider}, {Seager}, {Traub}, \&
  {Woolf}}]{DesMarais2002}
{Des Marais}, D.~J., {Harwit}, M.~O., {Jucks}, K.~W., {et~al.} 2002,
  Astrobiology, 2, 153

\bibitem[{{Fischer} {et~al.}(2016){Fischer}, {Anglada-Escude}, {Arriagada},
  {Baluev}, {Bean}, {Bouchy}, {Buchhave}, {Carroll}, {Chakraborty}, {Crepp},
  {Dawson}, {Diddams}, {Dumusque}, {Eastman}, {Endl}, {Figueira}, {Ford},
  {Foreman-Mackey}, {Fournier}, {F{\H{u}}r{\'e}sz}, {Gaudi}, {Gregory},
  {Grundahl}, {Hatzes}, {H{\'e}brard}, {Herrero}, {Hogg}, {Howard}, {Johnson},
  {Jorden}, {Jurgenson}, {Latham}, {Laughlin}, {Loredo}, {Lovis}, {Mahadevan},
  {McCracken}, {Pepe}, {Perez}, {Phillips}, {Plavchan}, {Prato}, {Quirrenbach},
  {Reiners}, {Robertson}, {Santos}, {Sawyer}, {Segransan}, {Sozzetti},
  {Steinmetz}, {Szentgyorgyi}, {Udry}, {Valenti}, {Wang}, {Wittenmyer}, \&
  {Wright}}]{Fischer2016}
{Fischer}, D.~A., {Anglada-Escude}, G., {Arriagada}, P., {et~al.} 2016, \pasp,
  128, 066001

\bibitem[{{Fujii} {et~al.}(2018){Fujii}, {Angerhausen}, {Deitrick},
  {Domagal-Goldman}, {Grenfell}, {Hori}, {Kane}, {Pall{\'e}}, {Rauer},
  {Siegler}, {Stapelfeldt}, \& {Stevenson}}]{Fujii2018}
{Fujii}, Y., {Angerhausen}, D., {Deitrick}, R., {et~al.} 2018, Astrobiology,
  18, 739

\bibitem[{{Gordon} {et~al.}(2017){Gordon}, {Rothman}, {Hill}, {Kochanov},
  {Tan}, {Bernath}, {Birk}, {Boudon}, {Campargue}, {Chance}, {Drouin}, {Flaud},
  {Gamache}, {Hodges}, {Jacquemart}, {Perevalov}, {Perrin}, {Shine}, {Smith},
  {Tennyson}, {Toon}, {Tran}, {Tyuterev}, {Barbe}, {Cs{\'a}sz{\'a}r}, {Devi},
  {Furtenbacher}, {Harrison}, {Hartmann}, {Jolly}, {Johnson}, {Karman},
  {Kleiner}, {Kyuberis}, {Loos}, {Lyulin}, {Massie}, {Mikhailenko},
  {Moazzen-Ahmadi}, {M{\"u}ller}, {Naumenko}, {Nikitin}, {Polyansky}, {Rey},
  {Rotger}, {Sharpe}, {Sung}, {Starikova}, {Tashkun}, {Auwera}, {Wagner},
  {Wilzewski}, {Wcis{\l}o}, {Yu}, \& {Zak}}]{Gordon2017hitran}
{Gordon}, I.~E., {Rothman}, L.~S., {Hill}, C., {et~al.} 2017, \jqsrt, 203, 3

\bibitem[{{Hegde} \& {Kaltenegger}(2013)}]{Hegde2013}
{Hegde}, S., \& {Kaltenegger}, L. 2013, Astrobiology, 13, 47

\bibitem[{{Johns} {et~al.}(2018){Johns}, {Marti}, {Huff}, {McCann},
  {Wittenmyer}, {Horner}, \& {Wright}}]{Johns2018}
{Johns}, D., {Marti}, C., {Huff}, M., {et~al.} 2018, \apjs, 239, 14

\bibitem[{{Kaltenegger}(2017)}]{Kaltenegger2017}
{Kaltenegger}, L. 2017, Annual Review of Astronomy and Astrophysics, 55, 433

\bibitem[{{Kaltenegger} \& {Traub}(2009)}]{KalteneggerTraub2009}
{Kaltenegger}, L., \& {Traub}, W.~A. 2009, The Astrophysical Journal, 698, 519

\bibitem[{{Kaltenegger} {et~al.}(2007){Kaltenegger}, {Traub}, \&
  {Jucks}}]{Kaltenegger2007}
{Kaltenegger}, L., {Traub}, W.~A., \& {Jucks}, K.~W. 2007, The Astrophysical
  Journal, 658, 598

\bibitem[{{Kane} {et~al.}(2016){Kane}, {Hill}, {Kasting}, {Kopparapu},
  {Quintana}, {Barclay}, {Batalha}, {Borucki}, {Ciardi}, {Haghighipour},
  {Hinkel}, {Kaltenegger}, {Selsis}, \& {Torres}}]{Kane2016}
{Kane}, S.~R., {Hill}, M.~L., {Kasting}, J.~F., {et~al.} 2016, \apj, 830, 1

\bibitem[{{Kasting} {et~al.}(1993){Kasting}, {Whitmire}, \&
  {Reynolds}}]{Kasting1993}
{Kasting}, J.~F., {Whitmire}, D.~P., \& {Reynolds}, R.~T. 1993, \icarus, 101,
  108

\bibitem[{King {et~al.}(1997)King, Tsay, Platnick, Wang, \& Liou}]{King1997}
King, M., Tsay, S.-C., Platnick, S., Wang, M., \& Liou, K. 1997, MODIS
  Algorithm Theoretical Basis Document, No. ATBD-MOD-05

\bibitem[{Kokaly {et~al.}(2017)Kokaly, Clark, Swayze, Livo, Hoefen, Pearson,
  Wise, Benzel, Lowers, Driscoll, {et~al.}}]{Kokaly2017usgs}
Kokaly, R.~F., Clark, R.~N., Swayze, G.~A., {et~al.} 2017, USGS spectral
  library version 7, Tech. rep., US Geological Survey

\bibitem[{Krissansen-Totton {et~al.}(2016)Krissansen-Totton, Schwieterman,
  Charnay, Arney, Robinson, Meadows, \& Catling}]{Krissansen2016}
Krissansen-Totton, J., Schwieterman, E.~W., Charnay, B., {et~al.} 2016, ApJ,
  817, 31

\bibitem[{{Lederberg}(1965)}]{Lederberg1965}
{Lederberg}, J. 1965, \nat, 207, 9

\bibitem[{{Lopez-Morales} {et~al.}(2019){Lopez-Morales}, {Currie}, {Teske},
  {Gaidos}, {Kempton}, {Males}, {Lewis}, {Rackham}, {Ben-Ami}, {Birkby},
  {Charbonneau}, {Close}, {Crane}, {Dressing}, {Froning}, {Hasegawa},
  {Konopacky}, {Kopparapu}, {Mawet}, {Mennesson}, {Ramirez}, {Stelter},
  {Szentgyorgyi}, {Wang}, {Alam}, {Collins}, {Dupree}, {Karovska}, {Kirk},
  {Levi}, {McGruder}, {Packman}, {Rugheimer}, \& {Rukdee}}]{LopezMorales2019}
{Lopez-Morales}, M., {Currie}, T., {Teske}, J., {et~al.} 2019, \baas, 51, 162

\bibitem[{{Lovelock}(1965)}]{Lovelock1965}
{Lovelock}, J.~E. 1965, \nat, 207, 568

\bibitem[{Lundock {et~al.}(2009)Lundock, Ichikawa, Okita, Kurita, Kawabata,
  Uemura, Yamashita, Ohsugi, Sato, \& Kino}]{Lundock2009}
Lundock, R., Ichikawa, T., Okita, H., {et~al.} 2009, A{\&}A, 507, 1649

\bibitem[{{Madden} \& {Kaltenegger}(2020)}]{Madden2020mnras}
{Madden}, J., \& {Kaltenegger}, L. 2020, \mnras, doi:10.1093/mnras/staa387

\bibitem[{{Madden} \& {Kaltenegger}(2018)}]{Madden2018Abio}
{Madden}, J.~H., \& {Kaltenegger}, L. 2018, Astrobiology, 18, 1559

\bibitem[{{Niro} {et~al.}(2005){Niro}, {Jucks}, \& {Hartmann}}]{Niro2005}
{Niro}, F., {Jucks}, K., \& {Hartmann}, J.~M. 2005, \jqsrt, 95, 469

\bibitem[{{O'Malley-James} \& {Kaltenegger}(2018)}]{OMalleyJames2018fluor}
{O'Malley-James}, J.~T., \& {Kaltenegger}, L. 2018, \mnras, 481, 2487

\bibitem[{{O'Malley-James} \& {Kaltenegger}(2019)}]{OMalleyJames2019}
---. 2019, \apjl, 879, L20

\bibitem[{{Ramsay} {et~al.}(2020){Ramsay}, {Amico}, {Bezawada}, {Cirasuolo},
  {Derie}, {Egner}, {George}, {Gont{\'e}}, {Gonz{\'a}lez Herrera},
  {Hammersley}, {Haupt}, {Heijmans}, {Ives}, {Jakob}, {Kerber}, {Koehler},
  {Mainieri}, {Manescau}, {Oberti}, {Padovani}, {Peroux}, {Siebenmorgen},
  {Tamai}, \& {Vernet}}]{Ramsay2020}
{Ramsay}, S., {Amico}, P., {Bezawada}, N., {et~al.} 2020, in Society of
  Photo-Optical Instrumentation Engineers (SPIE) Conference Series, Vol. 11203,
  \procspie, 1120303

\bibitem[{{Rauer} {et~al.}(2011){Rauer}, {Gebauer}, {Paris}, {Cabrera},
  {Godolt}, {Grenfell}, {Belu}, {Selsis}, {Hedelt}, \& {Schreier}}]{Rauer2011}
{Rauer}, H., {Gebauer}, S., {Paris}, P.~V., {et~al.} 2011, \aap, 529, A8

\bibitem[{{Rodler} \& {L{\'o}pez-Morales}(2014)}]{Rodler2014}
{Rodler}, F., \& {L{\'o}pez-Morales}, M. 2014, \apj, 781, 54

\bibitem[{Rossow \& Schiffer(1999)}]{RossowSchiffer1999}
Rossow, W.~B., \& Schiffer, R.~A. 1999, Bull. Amer. Meteorol. Soc., 80, 2261

\bibitem[{{Rugheimer} {et~al.}(2015){Rugheimer}, {Kaltenegger}, {Segura},
  {Linsky}, \& {Mohanty}}]{Rugheimer2015Spectra}
{Rugheimer}, S., {Kaltenegger}, L., {Segura}, A., {Linsky}, J., \& {Mohanty},
  S. 2015, \apj, 809, 57

\bibitem[{{Rugheimer} {et~al.}(2013){Rugheimer}, {Kaltenegger}, {Zsom},
  {Segura}, \& {Sasselov}}]{Rugheimer2013}
{Rugheimer}, S., {Kaltenegger}, L., {Zsom}, A., {Segura}, A., \& {Sasselov}, D.
  2013, Astrobiology, 13, 251

\bibitem[{{Schwieterman} {et~al.}(2018){Schwieterman}, {Kiang}, {Parenteau},
  {Harman}, {DasSarma}, {Fisher}, {Arney}, {Hartnett}, {Reinhard}, {Olson},
  {Meadows}, {Cockell}, {Walker}, {Grenfell}, {Hegde}, {Rugheimer}, {Hu}, \&
  {Lyons}}]{Scheieterman2018}
{Schwieterman}, E.~W., {Kiang}, N.~Y., {Parenteau}, M.~N., {et~al.} 2018,
  Astrobiology, 18, 663

\bibitem[{{Segura} {et~al.}(2005){Segura}, {Kasting}, {Meadows}, {Cohen},
  {Scalo}, {Crisp}, {Butler}, \& {Tinetti}}]{Segura2005}
{Segura}, A., {Kasting}, J.~F., {Meadows}, V., {et~al.} 2005, Astrobiology, 5,
  706

\bibitem[{{Segura} {et~al.}(2003){Segura}, {Krelove}, {Kasting}, {Sommerlatt},
  {Meadows}, {Crisp}, {Cohen}, \& {Mlawer}}]{Segura2003}
{Segura}, A., {Krelove}, K., {Kasting}, J.~F., {et~al.} 2003, Astrobiology, 3,
  689

\bibitem[{{Shields} \& {Carns}(2018)}]{Shields2018}
{Shields}, A.~L., \& {Carns}, R.~C. 2018, \apj, 867, 11

\bibitem[{{Snellen} {et~al.}(2015){Snellen}, {de Kok}, {Birkby}, {Brandl},
  {Brogi}, {Keller}, {Kenworthy}, {Schwarz}, \& {Stuik}}]{Snellen2015}
{Snellen}, I., {de Kok}, R., {Birkby}, J.~L., {et~al.} 2015, \aap, 576, A59

\bibitem[{{Snellen} {et~al.}(2017){Snellen}, {D{\'e}sert}, {Waters},
  {Robinson}, {Meadows}, {van Dishoeck}, {Brand l}, {Henning}, {Bouwman},
  {Lahuis}, {Min}, {Lovis}, {Dominik}, {Van Eylen}, {Sing},
  {Anglada-Escud{\'e}}, {Birkby}, \& {Brogi}}]{Snellen2017}
{Snellen}, I.~A.~G., {D{\'e}sert}, J.~M., {Waters}, L.~B.~F.~M., {et~al.} 2017,
  \aj, 154, 77

\bibitem[{Traub(2003)}]{Traub2003}
Traub, W. 2003, in Astronomical Society of the Pacific Conference Series, Vol.
  294, Scientific Frontiers in Research on Extrasolar Planets, ed. D.~Deming \&
  S.~Seager, 595--602

\bibitem[{{Traub} \& {Jucks}(2002)}]{TraubJucks2002}
{Traub}, W.~A., \& {Jucks}, K.~W. 2002, Washington DC American Geophysical
  Union Geophysical Monograph Series, 130, 369

\bibitem[{{Traub} \& {Stier}(1976)}]{TraubStier1976}
{Traub}, W.~A., \& {Stier}, M.~T. 1976, \ao, 15, 364

\end{thebibliography}

\end{document}